\documentclass[fleqn,usenatbib, referee]{mnras}

\usepackage{newtxtext,newtxmath}

\usepackage[T1]{fontenc}
\usepackage{ae,aecompl}
\usepackage{gensymb}

\usepackage{graphicx}	
\usepackage{amsmath}	
\usepackage{amssymb}	

\title[A Glitch in SXP 1062]{Discovery of a Glitch in the Accretion Powered Pulsar SXP 1062}

\author[M. M. Serim, \c{S}. \c{S}ahiner, D. \c{C}erri--Serim, S. \c{C}. \.{I}nam and A. Baykal]
{M. M. Serim$^{1}$\thanks{E-mail: muhammed@astroa.physics.metu.edu.tr (MMS); seyda@astroa.physics.metu.edu.tr (\c{S}\c{S}); danjela@astroa.physics.metu.edu.tr (D\c{C}S); inam@baskent.edu.tr (S\c{C}\.{I}); altan@astroa.physics.metu.edu.tr (AB)}, \c{S}. \c{S}ahiner$^{1}$\footnotemark[1], D. \c{C}erri--Serim$^{1}$\footnotemark[1], S. \c{C}. \.{I}nam$^{2}$\footnotemark[1] and A. Baykal$^{1}$\footnotemark[1] \\
$^{1}$Physics Department, Middle East Technical University, 06531 Ankara, Turkey\\
$^{2}$Department of Electrical and Electronics Engineering, Ba\c{s}kent University, 06530 Ankara, Turkey}

\date{Accepted XXX. Received YYY; in original form ZZZ}

\pubyear{2017}

\begin{document}
\label{firstpage}
\pagerange{\pageref{firstpage}--\pageref{lastpage}}
\maketitle


\begin{abstract}

We present timing analysis of the accretion powered pulsar SXP 1062, based on the observations of \textit{Swift}, \textit{XMM-Newton} and \textit{Chandra} satellites covering a time span of about 2 years. We obtain a phase coherent timing solution which shows that SXP 1062 has been steadily spinning down with a rate $-\,4.29(7) \times 10^{-14}$ Hz s$^{-1}$ leading to a surface magnetic field estimate of about $1.5 \times 10^{14}$ G. We also resolve the binary orbital motion of the system from X-ray data which confirms an orbital period of 656(2) days. On MJD 56834.5, a sudden change in pulse frequency occurs with $\Delta\nu = 1.28(5) \times 10^{-6} $ Hz, which indicates a glitch event. The fractional size of the glitch is $\Delta \nu / \nu \! \sim \! 1.37(6) \times 10^{-3}$ and SXP 1062 continues to spin-down with a steady rate after the glitch. A short X-ray outburst 25 days prior to the glitch does not alter the spin-down of the source; therefore the glitch should be associated with the internal structure of the neutron star. While glitch events are common for isolated pulsars, the glitch of SXP 1062 is the first confirmation of the observability of this type of events among accretion powered pulsars. Furthermore, the value of the fractional change of pulse frequency ensures that we discover the largest glitch reported up to now. 

\end{abstract}


\begin{keywords}
X-rays: binaries -- pulsars: individual: SXP 1062 -- stars: neutron -- accretion, accretion discs
\end{keywords}



\section{Introduction}

SXP 1062 is a Be/X-ray binary (BeXRB) system discovered in the eastern wing of the Small Magellanic Cloud (SMC) \citep{henault2012}. The detection of strong X-ray pulsations with a period of 1062 s revealed that the compact object in this system, is a slowly rotating pulsar. Furthermore, SXP 1062 is associated with a supernova remnant (SNR) and its kinematic age is calculated to be as young as 10--40 kyr \citep{henault2012,haberl2012}. The theoretical contradiction between its long period and young age puts SXP 1062 into the center of a remarkable attention.

The spectral type of the optical counterpart 2dFS 3831 \citep{evans2004} is specified to be B0-0.5(III)e+ \citep{henault2012}. Moreover, spectroscopic observations imply that the large circumstellar disc around the counterpart has been growing in size or density \citep{sturm2013}. I-band photometry carried on by \textit{Optical Gravitational Lensing Experiment} \citep[OGLE;][]{udalskietal2008} revealed periodic magnitude variations which signify that the orbital period of the binary system is likely to be $\sim \! 656$ d \citep{schmidtke2012}.

The X-ray spectrum of SXP 1062 is basically described by an absorbed powerlaw. Additionally, thermal components are used to describe a possible soft excess below 1 keV \citep{henault2012,sturm2013}. A $3\sigma$ evidence for \textit{Fe} K$\alpha$ emission line at 6.4 keV is also reported by \cite{sturm2013}. Furthermore, a search for a proton cyclotron absorption line on the 0.2--10 keV continuum yields no significant evidence \citep{sturm2013}.

First four pulse period measurements of SXP 1062 in 2010, demonstrate a very high spin-down rate of $\dot{\nu}=-\,2.6 \times10^{-12}$ Hz s$^{-1}$ during a short observation interval of 18 days \citep{haberl2012}. The following period is measured $\sim \! 2.5$ years later, which implies a 40 factor lower long term spin-down rate \citep{sturm2013}. In view of the fact that most BeXRBs show spin-up during their outbursts \citep{bildsten1997}, SXP 1062 is atypical with its strong spin-down.

The extraordinary observational properties of SXP 1062; such as its long pulse period, strong spin-down and young age; have led several authors to implement different theoretical models in order to explain the true nature of the source \citep{oskinova2013}. First, \cite{haberl2012} suggested that the initial pulse period of the neutron star at birth might be unusually long and calculated a lower limit of 0.5 s. Another explanation given by \cite{popov2012} indicated that the initial magnetic field of the neutron star at birth might be as large as $10^{14}$ G which then experienced a field decay. On the other hand, \cite{ikhsanov2012} proposed that the initial magnetic field could be $\sim \! 4 \times 10^{13}$ G once the magnetization of the accretion flow is introduced within the scope of the magnetic accretion scenario. \cite{fu2012} pointed out that SXP 1062 may currently have a magnetic field higher than $10^{14}$ G, although no cyclotron line is detected on the X-ray spectrum. The possibility of high magnetic field nominates SXP 1062 as an accreting magnetar however, \cite{postnov2014} claim that quasi-spherical accretion theory \citep{shakura2012,shakura2013} estimates a lower limit for the magnetic field which is consistent with the standard values for neutron stars.

In this paper, we present timing analysis of the accretion powered pulsar SXP 1062 with \textit{Swift}, \textit{XMM-Newton} and \textit{Chandra} observations between 2012 Oct. 9 and 2014 Nov. 23. First, we find orbital parameters of the binary system. In addition, we resolve a glitch event on MJD 56834.5 which is the first observational evidence that accretion powered pulsars may glitch like isolated pulsars. Furthermore, the fractional size of the glitch ensures that we discover the largest glitch reported upto now. In Section 2, we describe the observations. Then, in Section 3, we explain the timing analysis and its results. Finally in Section 4, we discuss that the observed glitch should be associated with the internal structure of the neutron star.

\section{Observations}

\begin{figure*}
  \center{\includegraphics[width=9cm,angle=270]{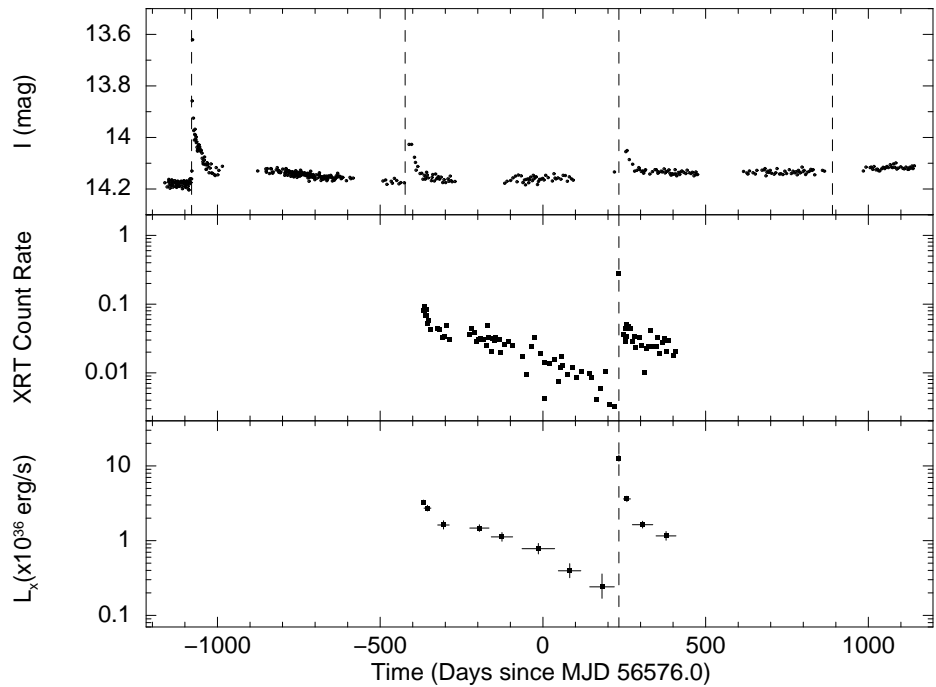}} 
  \caption{\textit{Upper panel:} I-band optical light curve of the counterpart provided by OGLE. The dashed lines indicate the times of optical outbursts calculated according to $\sim\!656$ d orbital period \citep{schmidtke2012}. \textit{Middle panel:} X-ray light curve of SXP 1062 from \textit{Swift}--XRT observations. Each point represents an observation. The dashed line indicates the time of the X-ray outburst, that is MJD 56809.5. \textit{Lower panel:} X-ray luminosity evolution in 0.3--10 keV energy band (calculated by using the distance to SMC: 60 kpc \citep{hilditch2005}). The maximum luminosity, $1.3 \times 10^{37}$ erg s$^{-1}$, is indicated with a dashed line.}
  \label{burst}
\end{figure*}

\textit{Swift} monitoring campaign of SXP 1062 begins during an X-ray outburst on 2012 Oct. 9 and continues until 2014 Nov. 23. Throughout a time about 2 years, 87 pointing observations have a total \textit{Swift}--XRT \citep[X-Ray Telescope;][]{burrows2005} exposure of $\sim \! 164$ ks. XRT operates in 0.2--10 keV energy range, with an effective area of 110 cm$^{2}$ at 1.5 keV. Its spatial resolution is 18 arcsec and its field of view (FOV) is 23.6 arcmin $\times$ 23.6 arcmin. This focusing instrument automatically switches between four operation modes depending on the source count rate. The operation mode of SXP 1062 observations is photon counting (PC) mode, which has a timing resolution of $\sim \! 2.5$ s. Clean event files are produced with the script \verb"XRTPIPELINE v.0.13.2" by using a default screening criteria. Light curves and spectra are extracted with \verb"XSELECT v.2.4d". Circular regions selected for source and background extraction have a radius of 35 and 141 arcsec, respectively. 

In order to investigate the evolution of X-ray luminosity, consecutive 8--10 observations are combined and spectral files are extracted for each group of observations. Spectra are re-binned to have at least 5 counts per bin and Cash statistic \citep{cash1979} is preferred during spectral fitting with \verb"XSPEC v.12.9.0". We model the spectra with an absorbed powerlaw and measure the X-ray flux in 0.3--10 keV energy band. Then, the flux values are converted to luminosity by using the distance to SMC, that is 60 kpc \citep{hilditch2005}. X-ray luminosity evolution is plotted on the lower panel of Figure \ref{burst}. 

OGLE \citep{udalskietal2008} monitoring of SXP 1062 provides I-band photometry measurements of the optical counterpart, which are received from the X-Ray variables OGLE Monitoring System \citep[XROM\footnote{http://ogle.astrouw.edu.pl/ogle4/xrom/xrom.html};][]{udalski2008}. Periodic optical outbursts are evident on the OGLE light curve (see the upper panel of Fig. \ref{burst}). This periodicity of $\sim \! 656$ d is associated with the binary orbit of the system \citep{schmidtke2012}.

SXP 1062 is also observed with \textit{XMM-Newton} and \textit{Chandra} observatories. The \textit{XMM-Newton} observation has a duration of $\sim \! 86$ ks on 2013 Oct. 11. We use single and double-pixel events (PATTERN 0--4) of EPIC-PN camera \citep{struder2001} which is sensitive to the photons in 0.15--15 keV energy range and possesses a FOV of 30 arcmin with a spatial resolution of 6 arcsec. During this observation, EPIC-PN camera operated on full frame mode which has a time resolution of 73 ms. The data reduction is carried out with \verb"SAS v.15.0.0" software. Filtering of high energy background flare times yields a net exposure of $\sim \! 47$ ks. We also avoid bad pixels by rejecting events with \verb"FLAG" $\neq 0$. In addition, 3 pointing observations taken with \textit{Chandra}-ACIS detectors \citep{garmire2003} are between 2014 June 29 and July 18, with a total exposure of $\sim \! 87$ ks. ACIS detectors have 17 arcmin $\times$ 17 arcmin FOV and operates in 0.2--10 keV energy band. All observations are conducted in imaging mode with a time resolution of $\sim \! 0.5$ s. The data is analysed with \verb"CIAO v.4.9" software. Default screening criteria are applied while producing the clean events. We extract 0.2--12 keV \textit{XMM-Newton} and 0.2--10 keV \textit{Chandra} lightcurves with 1 s bin time and all time series are converted to Solar System barycentre. Circular source extraction regions have radii of 25 and 8 arcsec for \textit{XMM-Newton} and \textit{Chandra}, respectively. Background emissions are estimated from source-free circular regions on the same detector chip as the source. 

\section{Timing Analysis}

\begin{figure*}
  \center{\includegraphics[width=9.6cm,angle=270]{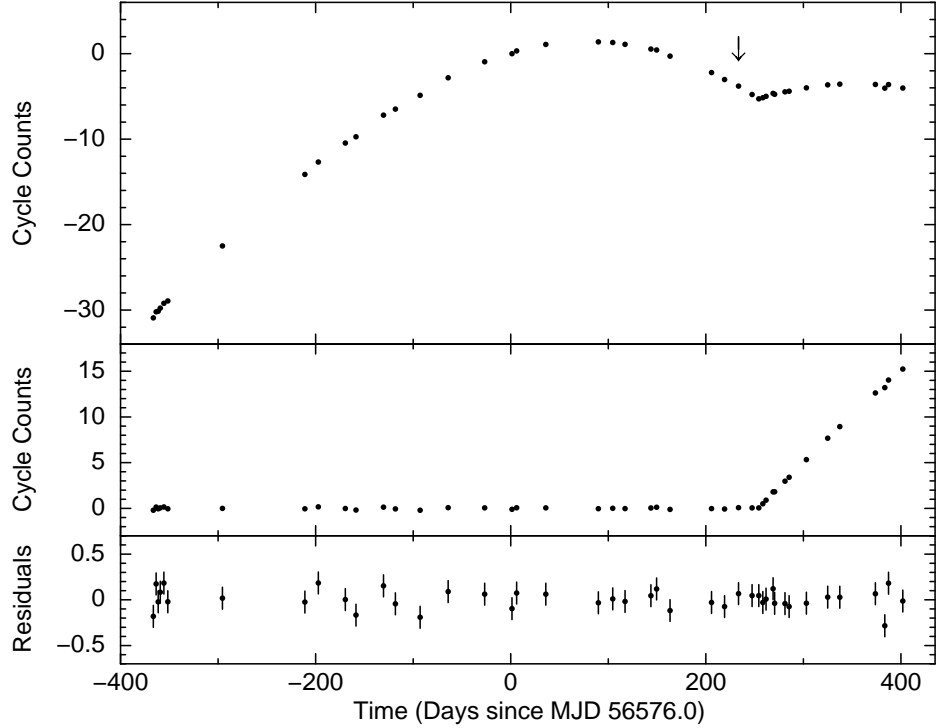}} 
  \caption{\textit{Upper panel:} Cycle counts of pulse phase obtained by expanding the TOAs. The time denoted with an arrow mark corresponds to the X-ray outburst of SXP 1062, that is MJD 56809.5. \textit{Middle panel:} Phase offset series after the removal of the secular spin-down trend and the orbital model given in Table \ref{soln}. The glitch event is evident on MJD 56834.5 and it is 25 days after the X-ray outburst. \textit{Lower panel:} Residuals after an additional removal of the glitch event. \textbf{The final model has a reduced $\chi^2$ of 1.0.}}
  \label{glitch}
\end{figure*}

\begin{table}
\caption{Binary orbit, timing solution and glitch parameters of SXP 1062. A number given in parentheses is the $1\sigma$ uncertainty in the least significant digit of a stated value.}
\label{soln}
 \center{\renewcommand{\arraystretch}{1.2}\begin{tabular}{lc}
  \hline 
  \textbf{Circular Orbital Model:} & \\
  Orbital Epoch (MJD) & 56351(10) \\
 $P_{\mathrm{orb}}$ (days) & 656(2) \\
 $\frac{a_{\mathrm{x}}}{c} \sin i $ (lt-s) & 1636(16) \\ 
 \hline
 \textbf{Timing Solution:} & \\
  Folding Epoch (MJD) & 56576.0 \\
  Validity Range (MJD) & 56209 -- 56830 \\
 $\nu_{\mathrm{o}}$ (mHz) & 0.931787(5) \\
 $\dot{\nu}_{\mathrm{o}}$ (Hz s$^{-1}$) & $-\,4.29(7) \times 10^{-14}$ \\
 \hline
 \textbf{Glitch Parameters:} & \\
 $t_\mathrm{g}$ (MJD) & $\sim \! 56834.5$ \\
 Validity Range (MJD) & 56834 -- 56978 \\
 $\Delta\nu$ (Hz) & $1.28(5) \times 10^{-6}$ \\
 $\Delta \dot{\nu}$ (Hz s$^{-1}$) & $- \, 1.5(9) \times 10^{-14}$ \\
 $\Delta\nu / \nu_{\mathrm{o}}$ & $1.37(6) \times 10^{-3}$ \\
 $\Delta \dot{\nu} / \dot{\nu}_{\mathrm{o}}$ & 0.3(2) \\
  \hline
 \end{tabular}}
\end{table}

We extract barycentric light curves from \textit{Swift}, \textit{Chandra} and \textit{XMM-Newton} observations described in the previous section. First, we search for periodicity on the light curve of the \textit{XMM-Newton} observation on MJD 56576.7 by folding it on trial periods \citep{leahy1983}. The period that give the maximum $\chi^2$ value is 1073.5 s. Then, all observations are folded with the same ephemeris and frequency. For each observation, pulse profiles with 10 phase bins are constructed. Observations which have an exposure less than the pulse period of SXP 1062 yield pulse profiles with zero count bins, therefore they are excluded during the timing analysis. The pulse profiles are described with harmonic representation as \citep{deeter1985}

\begin{equation}
 f(\phi) = F_{\mathrm{o}} + \sum_{k=1}^{n} F_{k} \, \cos k(\phi-\phi_{k}) \, ,
\end{equation}

\noindent where $n$ is the number of harmonics used for the representation of a pulse. Similarly, the template pulse profile is obtained from the longest observation ($\sim\!47$ ks) which is held on MJD 56576.7 with \textit{XMM-Newton}. The template pulse profile is represented as

\begin{equation}
 g(\phi) = G_{\mathrm{o}} + \sum_{k=1}^{n} G_{k} \, \cos k(\phi-\phi_{k}) \, .
\end{equation}

The time of arrival (TOA) of each pulse is estimated by searching the location of the maximum in the cross-correlation with template pulse \citep{deeter1985}\citep[see also][for applications]{icdem2012},

\begin{equation}
 \Delta\Phi = \dfrac{\sum\limits_{k=1}^{n} k \, G_{k} \, F_{k} \, \sin k \Delta\phi_{k}}{\sum\limits_{k=1}^{n}k^{2} \,
G_{k} \, F_{k} \, \cos k \Delta\phi_{k}} \,.
\end{equation}

\noindent To check how pulse timing is affected by the changes in pulse profiles, we measure TOAs by using different number of harmonics. The obtained TOAs are consistent within $1\sigma$ level for all harmonics therefore, we decide to perform pulse timing analysis by using five harmonics. In order to examine the timing behaviour of SXP 1062, we first focus on pre-outburst TOAs which are described as

\begin{equation}
 \Phi (t) = \Phi_{\mathrm{o}} + \nu_{\mathrm{o}} \, (t-t_{\mathrm{o}}) + \frac{1}{2} \, {\dot\nu_{\mathrm{o}}} \, (t-t_{\mathrm{o}})^{2} \, ,
\end{equation}

\noindent where $t_{\mathrm{o}}$ is the folding epoch, $\nu_{\mathrm{o}}$ is the spin frequency and ${\dot\nu_{\mathrm{o}}}$ is the derivative of the spin frequency, respectively. We expand TOAs around the spin-down rate previously reported by \cite{sturm2013} and obtain a phase coherent timing solution (see the upper panel of Fig. \ref{glitch}). We find that the source is spinning down with a rate of ${\dot\nu_{\mathrm{o}}}= -\,4.29(7) \times 10^{-14}$ Hz s$^{-1}$ between MJD 56209 and MJD 56830 (see Table \ref{soln}).

\begin{figure}
  \center{\includegraphics[width=5.1cm,angle=270]{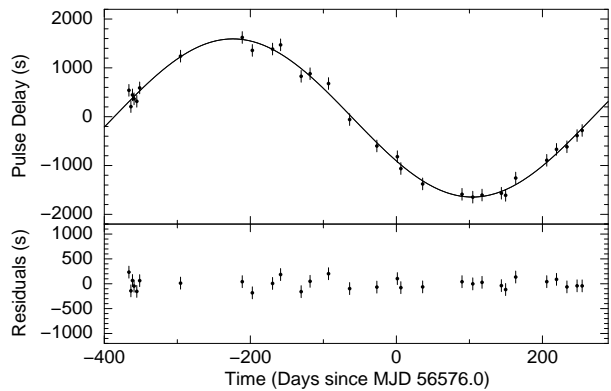}} 
  \caption{\textit{Upper panel:} Doppler delays on the residuals after the removal of the spin-down from TOAs prior to the glitch, are fitted with a circular orbital model (solid line) \textbf{with a reduced $\chi^2$ of 1.0}. The orbital parameters are given in Table \ref{soln}. \textit{Lower panel:} Residuals after the removal of orbital model.}
  \label{orbit}
\end{figure}

The residuals after the removal of the spin-down trend are given in Figure \ref{orbit}. The fluctuation on the residuals is consistent with Doppler delays due to orbital motion. In general for an eccentric orbit, it can be represented as \citep{deeter1981}\citep[see also][for applications]{zand2001,baykal2000,baykal2010}

\begin{equation}
\begin{split}
 \delta t_{\mathrm{orbit}} = \, & x \, \sin (l) - \frac{3}{2} \, x \, e \sin (w) \\
 &+ \frac{1}{2} \, x \, e \cos (w) \sin (2l) - \frac{1}{2} \, x \, e \sin (w) \cos (2l) \, ,
 \end{split}
 \label{eqn:orbit}
 \end{equation}

\noindent where $x \!=\! \frac{a_{\mathrm{x}}}{c}\sin i$ is the light travel time for projected semi-major axis, $i$ is the inclination angle between the line of sight and the orbital angular momentum vector, $l \!=\! 2 \pi \, (t - T_{\mathrm{\frac{\pi}{2}}}) \, / \, P_{\mathrm{orb}} + \frac{\pi}{2}$ is the mean orbital longitude at $t$, $T_{\mathrm{\frac{\pi}{2}}}$ is the epoch when the mean orbital longitude is equal to $90\degree$, $P_{\mathrm{orb}}$ is the orbital period, $w$ is the longitude of periastron and $e$ is the eccentricity. Since the time span of fitted TOAs covers less than one orbital cycle, we fix the orbital period to $P_{\mathrm{orb}} \!=\! 656(2)$ d as reported by \cite{schmidtke2012} and seek for other orbital parameters. 

\begin{figure}
  \center{\includegraphics[width=5.8cm,angle=270]{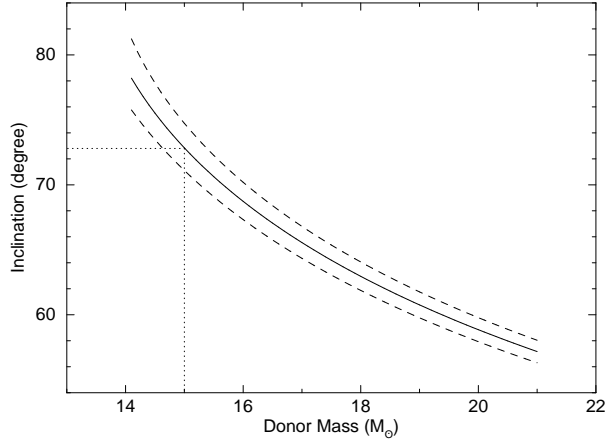}} 
  \caption{The relation between the binary system inclination angle and the donor mass in units of M$_{\odot}$. The solid curve is constructed by using the mass function value of 10.9M$_{\odot}$, whereas the dashed curves are constructed for 1$\sigma$ error in the mass function ($\pm 0.3$M$_{\odot}$). For a donor mass of $ \sim \! 15$M$_{\odot}$ \citep{henault2012}, we find that the inclination angle would be $\simeq\! 73(2) \degree$.}
  \label{inclination}
\end{figure}

The best fit with a reduced $\chi^2$ of 1.0 (see Fig. \ref{orbit} and Table \ref{soln}) yields a circular orbit with orbital parameters $\frac{a_{\mathrm{x}}}{c} \sin i \!=\! 1636(16)$ lt-s and $T_{\mathrm{\frac{\pi}{2}}} \!=\! 56351(10)$ MJD. 
Furthermore, a search for eccentricity leaving all orbital fit parameters free including the orbital period, gives an estimate for the upper limit to the eccentricity as 0.2 at 1.6$\sigma$ that is 90 per cent confidence level \citep{lampton1976}. Using these results, the mass function $f(M)$ can be calculated via

\begin{equation}
 f(M) = \frac{4\pi^{2}}{G} \, \frac{(a_{\mathrm{x}} \sin i)^{3}}{P_{\mathrm{orb}}^{2}} = \frac{(M_{\mathrm{c}} \sin i)^{3}}{(M_{\mathrm{x}} \! + \! M_{\mathrm{c}})^{2}} \, ,
\end{equation}

\noindent where $M_{\mathrm{c}}$ is the mass of the counterpart and $M_{\mathrm{x}}$ is the mass of the neutron star. The mass function calculated via orbital parameters is $f(M) \!\simeq\! 10.9(3)$M$_{\odot}$. Assuming a neutron star with a mass of 1.4M$_{\odot}$, we plot the relation between the inclination angle and the donor mass in Figure \ref{inclination}. Considering the typical evolutionary mass of the optical counterpart as $ \sim \! 15$M$_{\odot}$ \citep{henault2012}, the inclination angle would be $i \!\simeq\! 73(2) \degree$.

In the middle panel of Figure \ref{glitch}, we show residuals after the removal of both the spin-down trend and the orbital model from all TOAs. The TOAs after MJD 56834.5 show a sudden change of slope which indicates a possible glitch event. We model TOAs after the glitch with a second order polynomial $ \Delta\phi = \phi_{\mathrm{o}} + \delta\nu \, (t-t_{\mathrm{g}}) +\delta \dot{\nu} \, (t-t_{\mathrm{g}})^2$, where $t_{\mathrm{g}}$ is the time of the glitch event. 
Thus, pulse frequencies receive a correction of $\nu_{\mathrm{o}} +\delta\nu +\delta\dot{\nu} \, (t-t_{\mathrm{g}})$. We find that the glitch event occurs on MJD 56834.5 and it causes a frequency shift of $\Delta\nu = 1.28(5) \times 10^{-6} $ Hz along with a change of spin-down rate $\Delta \dot{\nu} = - \, 1.5(9) \times 10^{-14}$ Hz s$^{-1}$ (see Table \ref{soln}). In the lower panel of Figure \ref{glitch}, we show residuals after the glitch correction.

\section{Discussion}

\subsection{The Orbit}

The accretion powered pulsar SXP 1062 is in a BeXRB system. In BeXRBs, the neutron star accretes matter from the stellar wind of a main sequence star, which is in the form of a circumstellar disk around the mass donor. Generally, the orbits of BeXRBs are relatively wide and moderately eccentric \citep[$P_\mathrm{orb} \! \geq \! 20$ d, $ e \! \geq \! 0.3$;][]{reig2011}. While the neutron star passes through the edge of circumstellar disk of the Be companion, it interacts with the material and accretion takes place. The X-ray emission of most BeXRBs is transient, however persistent BeXRBs with low luminosities ($L_\mathrm{x} \! \sim \! 10^{34-35}$ erg s$^{-1}$) also exist. Persistent systems are known to contain slower pulsars that have wider orbits \citep[$P_\mathrm{s} \! > \! 200$ s, $P_\mathrm{orb} \! > \! 200$ d;][]{reig1999,reig2011}. Although most BeXRBs have eccentric orbits, there are several systems (i.e. X Per, GS 0834--430, KS 1947+300) with low eccentricity ($e \! < \! 0.2$). 

Transient BeXRBS show two types of X-ray outbursts \citep{stella1986}. Type I outbursts ($L_\mathrm{x} \! \sim \! 10^{36-37}$ erg s$^{-1}$) occur once in a while the pulsar passes from the periastron of the orbit, where accretion enhances. On the other hand, Type II outbursts are major events ($L_\mathrm{x} \! \geq \! 10^{37}$ erg s$^{-1}$) that occur during mass ejection episodes of the counterpart. Transient BeXRBs may have very low quiescence luminosities ($L_\mathrm{x} \! \leq \! 10^{33}$ erg s$^{-1}$). The luminosity increase during a Type II outburst can be 3--4 order of magnitudes, while it is only about one order of magnitude for a Type I outburst \citep{reig2011}. When an X-ray outburst occurs in a BeXRB, the neutron star generally enters a spin-up episode due to enhanced accretion. 

As the population of BeXRBs has grown in the past decades, the pulse periods ($P_\mathrm{s}$) of pulsars in BeXRBs are still strongly correlated with the orbital periods ($P_\mathrm{orb}$), as it is firstly demonstrated by \cite{corbet1984}. Although there is a large scatter in data, a positive correlation is evident with $P_\mathrm{s} \! \propto \! P_\mathrm{orb}^2$. The large BeXRB population in SMC also obeys this relation \citep{knigge2011,yang2017}. Therefore in BeXRBs, pulsars with longer pulse periods reside in binary systems with wider orbits and consequently lower accretion rates.

The optical light curve of SXP 1062 (see Fig. \ref{burst}) shows periodic variations devoted to an orbital period of $\sim \! 656$ days \citep{schmidtke2012}. X-ray observations of the source reveal the occurrence of a Type I outburst ($L_\mathrm{x} \! \! \simeq \! 1.3 \times 10^{37}$ erg s$^{-1}$) that happen together with the optical enhancement. However, the long term spin-down of SXP 1062 is not interrupted by the outburst. Prior to the X-ray outburst a minimum luminosity of $2.4 \times 10^{35}$ erg s$^{-1}$ is measured, hence the luminosity increases by a factor of $\sim \! 50$ during the outburst. The outburst of SXP 1062 is observed only during an observation with an exposure of 2.2 ks, however the actual duration of the outburst might be longer since neighbouring observations are 14 days apart. The luminosity drops to $3.6 \times 10^{36}$ erg s$^{-1}$ therefore, the outburst finishes in the following observation. These characteristics classify SXP 1062 as a persistent BeXRB.

We are able to resolve the orbital motion of SXP 1062 from its X-ray emission observed for $ \sim \! 2$ years (see Fig. \ref{orbit}). We determine the orbital epoch as 56351(10) MJD and the light travel time for projected semi-major axis as 1636(16) lt-s by considering a circular orbit with a period of 656(2) days. (see Table \ref{soln}). We also report an upper limit of 0.2 to the eccentricity at 90 per cent confidence level, therefore SXP 1062 is claimed to be in a low eccentric orbit despite the fact that denser observational coverage is needed for a better assessment. The orbital and pulse periods of the system, position the source on a place in line with BeXRBs on the Corbet diagram, the uppermost right end of the existing correlation for BeXRBs. Moreover, the mass function of the system is calculated to be $f(M) \!\simeq\! 10.9(3)$M$_{\odot}$, which seems appropriate bearing in mind that the Be companion is suggested to have a mass of $ \sim \! 15$M$_{\odot}$ \citep{henault2012}. Consequently, the orbital inclination can be evaluated as $i \!\simeq\! 73(2) \degree$. If we allow variation of the donor mass, the effect on the inclination angle is plotted in Figure \ref{inclination}. Using this relation, the minimum donor mass is determined to be 13.3(3)M$_{\odot}$ for $i \!=\! 90 \degree$.

\subsection{Magnetic Field Estimation from Secular Spin-down Trend before the Glitch}

If we consider that the source is accreting via a prograde accretion disc that is formed before the glitch, standard accretion disc theory \citep{pringle1972,lamb1973,ghosh1979,wang1987,ghosh1994,torkelsson1998} can be used to estimate the surface magnetic field of the neutron star. For this scenario, the inner radius of the accretion disc at which the magnetosphere disrupts the Keplerian rotation depends on the accretion rate ($\dot{M}$) and the magnetic dipole moment of the neutron star ($\mu \!=\! BR^3$ where $B$ is the surface magnetic field and $R$ is the radius of the neutron star) as

\begin{equation}
r_\mathrm{o} = K \, \mu^{4/7} \, (GM)^{-1/7} \, \dot{M}^{-2/7} \, ,
\label{eqn:r0}
\end{equation}

\noindent where $K$ is a dimensionless parameter of about 0.91 and $M$ is the mass of the neutron star \citep{pringle1972,lamb1973}. The torque can then be estimated as

\begin{equation}
2\pi \, I \, \dot{\nu }= n(\omega_\mathrm{s}) \, \dot{M} \, l_\mathrm{K} \, ,
\label{eqn:torque}
\end{equation} 

\noindent where $I$ is the moment of inertia of the neutron star, $\dot{\nu}$ is the spin rate of the neutron star, $n(\omega_\mathrm{s})$ is the dimensionless torque which is a factor parametrising the material torque and magnetic torque contributions to the total torque, and $l_\mathrm{K }= (GMr_\mathrm{o})^{1/2}$ is the angular momentum per mass added by the Keplerian disc at $r_\mathrm{o}$. The dimensionless torque can be approximated as

\begin{equation}
n(\omega_\mathrm{s}) \approx 1.4 \, (1-\omega_\mathrm{s}/\omega_\mathrm{c}) \, / \, (1-\omega_\mathrm{s}) \, ,
\label{eqn:dimtorque}
\end{equation}

\noindent where $\omega_\mathrm{s}$, being equal to the ratio of the neutron star's rotational frequency to the Keplerian frequency at the inner radius of the accretion disc, is known as the fastness parameter and can be expressed as

\begin{equation}
\omega_\mathrm{s} = 2\pi \, K^{3/2} \, P^{-1} \, (GM)^{-5/7} \, \mu^{6/7} \, \dot{M}^{-3/7} \, ,
\label{eqn:fastness}
\end{equation}

\noindent where $P$ is the pulse period of the neutron star. In Eqn. \ref{eqn:dimtorque}, $\omega_\mathrm{c}$ is the critical fastness parameter which has been estimated to be $\sim \!0.35$ \citep{ghosh1979,wang1987,ghosh1994,torkelsson1998}. For $\omega_\mathrm{s} \!=\! \omega_\mathrm{c}$, the total torque on the neutron star becomes zero (i.e. $n(\omega_\mathrm{s}) \!=\! 0$) due to the negative torque contribution coming from the magnetic torque exerted outside the co-rotation radius at which the neutron star's rotational frequency equals to the Keplerian frequency.

For $\omega_\mathrm{s} \!>\! \omega_\mathrm{c}$, spin-down contribution coming from the outer disc outside the co-rotation radius is greater in magnitude than the total spin-up contributions coming from the material torque at the inner radius and the magnetic torque inside the co-rotation radius. This leads to a net spin-down of the neutron star (i.e. $n(\omega_\mathrm{s}) \!<\! 0$). On the contrary, for $\omega_\mathrm{s} \!<\! \omega_\mathrm{c}$, spin-up contribution coming from the material and magnetic torques is greater in magnitude than the spin-down contribution coming from the magnetic torques from the outer disc (i.e. $n(\omega_\mathrm{s}) \!>\! 0$). 

From a quadratic fit to the arrival times prior to the glitch, SXP 1062 is found to show a secular spin-down with a rate of $-\,4.29(7) \times 10^{-14}$ Hz s$^{-1}$ when a maximum luminosity of $L_\mathrm{x} \!\sim \! 3.3 \times 10^{36}$ erg s$^{-1}$ is observed. Considering this luminosity value to be nearly the total accretion luminosity (i.e. $L \!=\! GM \dot{M} / R$) and assuming a typical neutron star with $I \!=\! 10^{45}$ g cm$^2$, $M \!=\! 1.4$M$_{\odot}$ and $R \!=\! 10^6$ cm; Eqn.s \ref{eqn:r0} - \ref{eqn:fastness} are solved numerically to obtain $\mu$ of about $1.5 \times 10^{32}$ G cm$^{3}$ leading to a magnetic field estimate of about $1.5 \times 10^{14}$ G with $n(\omega_\mathrm{s}) \!\approx \! -\,0.0123$ and $r_\mathrm{o} \!=\! 8.78 \times 10^9$ cm. 

SXP 1062 can be considered to be a member of a class of accretion powered pulsars in high-mass X-ray binaries with very slow pulsations and persistent spin-down states \citep{reig2012,fu2012}. Long spin periods together with the spin-down behaviour of these pulsars are argued as an indication of their magnetar-like magnetic fields. Thus, this class is sometimes classified as ``accreting magnetars''. Alternatively by using a theoretical model based on quasi-spherical subsonic accretion, long spin periods of these systems have also been considered not to be necessarily related to magnetar fields \citep{shakura2013}. 

\cite{fu2012} previously made use of three different theoretical approaches to obtain an estimate of the magnetic field of SXP 1062: Firstly, they estimated the magnetic field strength by considering the time scale for the ejector phase being comparable to the estimated age of the pulsar. Secondly, they estimated the magnetic field strength assuming the short-term spin-down rate of $-2.6\times 10^{-12}$ Hz s$^{-1}$ as being near the maximum spin-down rate in disk or spherical accretion \citep{lynden1974,lipunov1982,bisnovatyi1991}. Their final approach was to make use of the spin-down mechanism proposed by \cite{illarionov1990}. All these three approaches lead to a surface magnetic field of SXP 1062 as $\gtrsim 10^{14}$ Gauss.

Our timing analysis shows that the source has a long-term secular steady spin-down trend with a rate of $-4.29(1)\times 10^{-14}$ Hz s$^{-1}$ which could be as a result of a steady disc accretion. Thus, using standard accretion theory, our magnetic field estimate for SXP 1062 follows consideration of accretion via prograde accretion disc with a small negative dimensionless torque. According to this theoretical framework, observed spin-down rate and luminosity of the source leads to a magnetar-like surface magnetic field estimation which is consistent with the previous estimations by \cite{fu2012}. 

\subsection{The Glitch}

\begin{figure}
  \center{\includegraphics[width=4.1cm,angle=270]{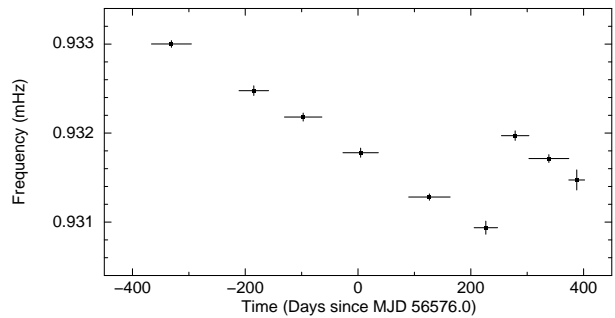}} 
  \caption{Pulse frequency evolution of SXP 1062. Frequencies are calculated from the slopes of linear fits to the TOAs shown in Fig. \ref{glitch}. The time intervals of linear fits are represented as x-axis error bars. The frequency jump on MJD 56834.5 is identified as a spin-up glitch event with $\Delta \nu = 1.28(5) \times 10^{-6}$ Hz. The source continues to spin-down after the glitch with a change of frequency derivative $\Delta \dot \nu = 1.5(9) \times 10^{-14}$ Hz s$^{-1}$.}
  \label{freqs}
\end{figure}

A glitch in the pulse frequency is observed 25 days after the X-ray outburst of SXP 1062. The source has not shown any spin-up trend during the outburst which may be due to a very short duration of the outburst. Actually, the outburst is displayed only in one of the observations, which has an exposure of about 2.2 ks. As seen from Figure \ref{glitch} and Table \ref{soln}, the glitch occurred on MJD 56834.5 with a change of pulse frequency $\Delta \nu = 1.28(5) \times 10^{-6}$ Hz and a change of pulse frequency derivative $\Delta \dot \nu = - \,1.5(9) \times 10^{-14}$ Hz s$^{-1}$. In Figure \ref{freqs}, we also show the pulse frequency evolution which is constructed by measuring the slopes of the TOAs (see Fig. \ref{glitch}) for time intervals of approximately 30--70 days. Since the occurrence of the glitch does not coincide with the time of the X-ray outburst, it should be associated with the internal structure of the neutron star. SXP 1062 continues to spin-down with a constant rate after the glitch event.

A glitch is a sudden fractional change in frequency which is mostly pursued by a change of spin-down rate of a previously rather stable rotating pulsar. Almost 10 per cent of pulsars are observed to glitch and pulsars of all ages seem to have glitches \citep{haskell2015} with fractional change of frequency ($\Delta \nu/\nu$) ranging from $10^{-11}$ to $10^{-5}$ and fractional change of frequency derivative ($\Delta\dot{\nu}/\dot{\nu}$) varying between $10^{-4}$ and $10^{-1}$ \citep{espinoza2011,yu2013,dib2014}. The core of a neutron star contains a significant amount of neutron superfluid \citep{lamb1978a,lamb1978b,sauls1989,lamb1991,datta1993,lattimer2007} therefore, the moment of inertia of the star resides mainly in the core. Moreover, the inner part of the crust lattice also contains an amount of neutron superfluid which carries $10^{-2}$ of the star's moment of inertia. Coupling time scales between crustal neutron superfluid and the rest of the crust is typically very long extending from months to years \citep{alpar1981,alpar1993,akbal2015}. For radio pulsars which spin-down due to electromagnetic dipole radiation, it is possible to resolve moment of inertia of the crustal superfluid during the post glitch \citep{espinoza2011,yu2013}. Like canonical pulsars, magnetars also exhibit glitches however, there are some distinguishing characteristics between these two groups. While almost all pulsar are radiatively quiet i.e. they are not accompanied by any burst or pulse profile changes after the glitch \citep{espinoza2011,yu2013} \citep[see][for exceptions]{archibald2016,manchester2011,livingstone2010}, magnetar glitches can either be radiatively loud i.e. they can be accompanied by flares, bursts and/or pulse profile changes; or radiatively quiet \citep{dib2014}.  

In magnetars, glitches are resolved with high fractional frequency changes at the order of $\Delta \nu / \nu \! \sim \! + \, 10^{-5}$  and $- \, 10^{-4}$ \citep{kaspi2017}. Largest spin-down glitches observed are, the glitch of 1E 2259+586 with $\Delta \nu / \nu \! \sim \! 10^{-6}$ \citep{archibald2013} and the glitch of SGR 1900+14 with $\Delta \nu / \nu \! \sim \! 10^{-4}$ within 80 days after a large outburst \citep{woods1999,thompson2000}. There are a few net spin-down glitches \citep{icdem2012,sasmaz2013,archibald2017} together with a large number of spin-up glitches \citep{dib2014}. Large spin-down glitches can be explained by particle outflow from magnetic multipoles during an outburst, while this process induces vortex inflow from the crust. The density of vortex lines are proportional to the superfluid velocity therefore the angular momentum taken from the crust \citep{thompson2000,duncan2013}. The spin-up glitches can be caused by sudden fractures of the crust and consequently vortex outflow in the crust superfluid \citep{thompson2000}. For both cases of spin-down and spin-up glitches, vortex unpinning from the crust occurs and then the vortices creep and re-pin to the crustal nuclei, therefore the post glitch relaxation should be observed in both cases \citep{gugercinoglu2014}.

Due to the presence of dominant external torque noise, it is not easy to detect these types of glitches for accretion powered pulsars in X-ray binaries \citep{baykal1997}. However, for KS 1947+300 \cite{galloway2004} have discovered a spin-up glitch. KS 1947+300 was spinning up during this glitch, therefore the influence of the external torques is not clear yet; whether the glitch event is associated with internal or external torques. Recently, \cite{ducci2015} have suggested that both glitches and anti-glitches are possible for accretion powered X-ray pulsars, furthermore glitches of binary pulsars should have longer rise and recovery time scales compared to isolated pulsars since they have pulse periods longer than those of isolated ones. 

SXP 1062 is found to be spinning down secularly until MJD 56834, that is 25 days subsequent to the X-ray outburst. Then, the source showed a spin-up glitch with a fractional frequency change of $\Delta \nu / \nu \! \sim \! 1.37(6) \times 10^{-3}$ and a fractional change of frequency derivative $\Delta \dot \nu / \dot \nu \! \sim \! 0.3(2)$.

During the secular spin-down of SXP 1062, the spin-down rate is measured to be $- \, 4.29(7) \times 10^{-14}$ Hz s$^{-1}$. If we consider that the observed glitch is due to a torque reversal (i.e. consider it as a frequency jump due to accretion torque) with a similar magnitude of spin-up rate, upper limit for ${{\Delta \nu} / {\nu}}$ can be estimated to be about $\sim \! 1.5 \times 10^{-5}$ for a maximum of $\Delta t \! \sim \! 4$ days (the time interval between two neighboring observations around the frequency jump) which is two orders of magnitude smaller than the observed ${{\Delta \nu} / {\nu}}$ of the glitch. So, it is unlikely that the glitch is as a result of the accretion torques. Furthermore, the ratio of the core superfluid moment of inertia to the crust moment of inertia ($I_{\mathrm{s}} / I_{\mathrm{c}}$) should be at the order of $10^2$ \citep{baykal1991,baykal1997}. Therefore the glitch event in SXP 1062 should be associated with the internal structure of the neutron star.

Recently, \cite{ducci2015} discussed observability of glitches in accretion powered pulsars by using the ``snowplow'' model of \cite{pizzochero2011}. In the two component neutron star model \citep{baym1969}, a neutron star consists of two components: the normal component where charged particles (protons and electrons) co-rotate with the neutron star's magnetic field with moment of inertia $I_{\mathrm{c}}$ and the neutron superfluid with moment of inertia $I_{\mathrm{s}}$. The rotating superfluid (both in the core and inner crust) is considered to be an array of vortices which are pinned to the crustal lattice of ions. When the neutron star slows down, a rotational lag is developed between the vortices and the normal component. Eventually, vortices are unpinned and suddenly move out after a certain critical value of rotational lag, leading to a glitch. The time required to build a glitch is inversely proportional to the spin-down rate therefore, pulsars with higher spin-down rates are expected to glitch more often. Moreover, the coupling time scales between the crust and core are proportional to the pulse period as $\tau = 10-100 \, P_{\mathrm{s}}$ \citep{alpar1984b,alpar1988,sidery2009}. Since SXP 1062 has a long pulse period along with a strong spin-down rate, it is a good candidate for observing such glitches. In accretion powered pulsars, the time scales for both glitch rise and decay are suggested to be long therefore, a glitch would appear as a single jump in frequency leaving the spin down-rate almost unchanged \citep{ducci2015}. The jump in pulse frequency can be estimated via \citep{ducci2015}

\begin{equation}
 \Delta\nu \simeq 2\times 10^{-5} \, \frac{Q_{0.95} \, R_{6}^{2} \, f_{15}}{M_{1.4} \, [ \, 1-Q_{0.95} \, (1-Y_{0.05}) \, ]} \qquad \mathrm{Hz} \, \mathrm{s^{-1}} \, ,
\end{equation}

\noindent where $Q$ ($= I_{\mathrm{s}} / (I_{\mathrm{c}} + I_{\mathrm{s}})$) is the fraction of superfluid in the neutron star ($Q_{0.95}$ in units of 0.95), $Y$ is the fraction of vortices coupled to normal crust ($Y_{0.05}$ in units of 0.05) and $f$ is the pinning force ($f_{15}$ in units of $10^{15}$ dyn cm$^{-1}$). The parameter $Y$ represents short time dynamics and approaches to 1 for long time scales (steady state). Assuming a neutron star with a mass of 1.4M$_{\odot}$, a radius of 10 km, $f_{15} \!\simeq\! 1$ dyn cm$^{-1}$ and by using the $\Delta\nu$ value observed for SXP 1062; we find that for a superfluid fraction around 95 per cent the fraction of coupled vortices is around 78 per cent. 

Both glitch rise and decay times for SXP 1062 should be at the order of a day or less ($\tau = 10-100 \, P_{\mathrm{s}} \simeq 10^{4}-10^{5}$ s) however, the sampling of TOAs around the glitch is about 3--4 days. Therefore; we observe neither the rise nor the decay of the glitch, since the glitch rise and decay should have already finished within the observational gaps. Thus for SXP 1062, the observed step-like change in pulse frequency and its magnitude can be qualitatively explained by the model of \cite{ducci2015}.

SXP 1062 has a strong and steady spin-down rate among accretion powered X-ray pulsars. Moreover, SXP 1062 is associated with a young supernova remnant with an age of 10--40 kyr \citep{henault2012,haberl2012}, therefore it is a young pulsar spinning down very fast in the remnant. The detection rate of glitches are observed to be higher for younger pulsars \citep{espinoza2011} and long intervals of steady spin rates are expected to increase glitch possibility \citep{ducci2015}. Therefore, these unique properties of SXP 1062 allows the vortices to creep and pin to the crustal nuclei \citep{alpar1984a,alpar1984b}. Sudden unpinning of vortices may cause a large glitch event, which is observed in this case with $\Delta \nu / \nu \! \sim \! 10^{-3}$ being the largest value of fractional frequency jump reported as far. The fractional size of the glitch suggests that $I_{\mathrm{s}}/I_{\mathrm{c}}$ is around $10^2$ which corresponds to soft equation of state \citep{datta1993,delsate2016}. It is possible to observe a glitch in this source again. In addition, the long pulse period of SXP 1062 makes it possible to reveal glitch rise and crust core coupling time if future observations are sampled closely \citep{newton2015}. Future monitoring of this source with \textit{LOFT} and \textit{NICER} can reveal more information about the interior of the neutron star. 

\section*{Acknowledgements}

We acknowledge support from T\"{U}B\.{I}TAK, the Scientific and Technological Research Council of Turkey through the research project MFAG 114F345. We thank M. Ali Alpar for helpful comments. We also thank the anonymous referee for the valuable comments that helped to improve the manuscript.






\bibliography{sxp1062}

\bibliographystyle{mnras}







\bsp	
\label{lastpage}
\end{document}